\documentclass{article}

\usepackage[left=1.3in,top=1.3in,right=1.3in,bottom=1.3in,nohead]{geometry}

\usepackage{url,subfig,latexsym,amsmath,amssymb, amsfonts,enumerate,
  epsfig, graphics,times, amsthm}
  
\usepackage[sort,compress]{cite}

\title{Finding the positive feedback loops underlying multi-stationarity}

\author{Elisenda Feliu\footnote{Department of Mathematical Sciences, University of Copenhagen; {\tt efeliu@math.ku.dk}},  \ \ Carsten Wiuf\footnote{Department of Mathematical Sciences, University of Copenhagen; {\tt wiuf@math.ku.dk}}}

\usepackage{enumerate,url}

\usepackage[version=3]{mhchem}
\usepackage{color}

\newcommand{\R}{\mathbb{R}}
\newcommand{\nM}{\widetilde{M}}

\DeclareMathOperator*{\sign}{sign}
\DeclareMathOperator*{\im}{im}

\begin{document}

\maketitle

\begin{abstract} 
Bistability is ubiquitous in biological systems. For example, bistability is found in many reaction networks that involve the control and execution of important biological functions, such as  signalling processes. Positive feedback loops, composed of species and reactions, are necessary for bistability, and  generally for multi-stationarity, to occur. These loops are therefore often used to illustrate and  pinpoint the parts of a multi-stationary network that are relevant (`responsible') for the observed multi-stationarity. However positive feedback loops are generally abundant in reaction networks but not all of them are  important for subsequent interpretation of the network's dynamics.

We present an automated procedure to determine the relevant positive feedback loops of a multi-stationary reaction network. The procedure only reports the loops that are relevant for multi-stationarity (that is, when broken multi-stationarity disappears) and not all positive feedback loops of the network. We show that the relevant  positive feedback loops must be understood in the context of the network (one loop might be relevant for one network, but cannot create multi-stationarity in another). Finally, we demonstrate the procedure by applying it to several examples of signalling processes, including a ubiquitination and an apoptosis network, and to models extracted from the Biomodels database. 

We have developed and implemented an automated procedure to find relevant  positive feedback loops in  reaction networks. The results of the  procedure are useful for interpretation and summary of the network's dynamics.
\end{abstract}

\section*{Background}

Bistability, and  multi-stationarity in general, is ubiquitous in biological systems ranging from biochemical networks to epidemiological and eco-systems \cite{santos,Markevich-mapk,Nguyen:2011p975,Zhdanov:2009p965}.  It is considered an important biological mechanism for controlling cellular and bacterial behaviour and developmental processes in organisms, and it is closely linked to the idea of the cell as a decision making unit, where a continuous input is converted to an on/off response corresponding to two distinct states of the cell \cite{Legewie:2005hw,palani}.

The question of bistability therefore arises naturally in many contexts. Many studies aim to demonstrate  that  in a given  biochemical system, bistability can or cannot occur \cite{Liu:2011p964,Markevich-mapk,harrington-feliu,Nguyen:2011p975,Eissing2004}. This has created some interest in formal methods that connect the network structure to the dynamic behaviour of the system, see e.g.~\cite{wiuf-feliu,conradi-switch,craciun-feinbergI,Tyson,Rand,PerezMillan,joshi-shiu-II,feinberg-def0,angelisontag2}.

One  qualitative network feature has in particular been  linked to multi-stationarity, namely the existence of a \emph{positive feedback loop}. A positive feedback loop consists of a sequence of species such that each species affects the production of another species, either positively or negatively, and such that the number of negative influences is even. The idea of associating positive feedback loops with bistability goes back to Jacob and Monod  who introduced it in the context of gene regulatory networks  \cite{jacob-monod}. It was later formalised  by Thomas in the form of a conjecture \cite{Thomas}, which was finally proved by Soul\'e \cite{soule}, see also \cite{gouze,kaufman}.

Soul\'e considers  dynamical systems of the form 
\begin{equation}\label{eq:ode1}
 \dot{x} = f(x),\qquad x\in \Omega\subseteq \R^n,
 \end{equation}
where $x=x(t)$,  $x=(x_1,\dots,x_n)$ is the vector of species concentrations, $\dot{x}=dx/dt$ is the derivative of $x$ with respect to time $t$, and $f$ is the so-called species-formation rate function, which specifies the instantaneous change in the concentrations.

The work of Soul\'e is based on the so-called \emph{interaction graph} \cite{soule}. This graph encodes how the variation of one species concentration depends on the concentration of the other species. It is built from the Jacobian matrix $J_f(x^*)$ of $f$ evaluated at a point $x^*$, such that the  non-zero entries of $J_f(x^*)$ corresponds to directed edges of the graph and the signs of the entries are edge labels. 
Soul\'e's analysis is mainly informative when the signs of the entries  are independent of   $x^*$\!.  He proved that the existence of a positive feedback loop   in the interaction graph is a necessary condition for $f(x)$ to have multiple zeros. In other words, it is a necessary condition  for multi-stationarity to exist in the ODE system \eqref{eq:ode1}. 

 Soul\'e's approach  usually works well when modelling gene networks, but it is often useless when modelling enzymatic signalling networks. In this case, the interaction graph rarely has constant labels.  A related line of work, that remedies this short-coming and might be seen as a refinement of Soul\'e's work, is based on the so-called \emph{directed species-reaction graph} (DSR-graph) \cite{craciun-feinbergII,banaji-craciun1,banaji-craciun2,wiuf-feliu}. If $f$ in  \eqref{eq:ode1} is obtained from a reaction network, then it decomposes in the form
\begin{equation}\label{eq:ode2}
\dot{x}= f(x) = A v(x),
 \end{equation}
where $A$ is the stoichiometric matrix of the network and $v(x)$ the vector of reaction rates. The DSR-graph uses this particular structure.

The DSR-graph is a bipartite graph  with nodes labeled   by the species and the  reactions of the reaction network. There is a signed directed edge from a species to a reaction if the species concentration contributes either positively (positive edge label) or negatively (negative edge label) to the rate $v(x)$ of the reaction. There is an edge from a reaction to a species  if the stoichiometric coefficient of the species in the reaction is non-zero. In this case the edge is labelled with the sign of the stoichiometric coefficient. Compared to the interaction graph, the DSR-graph makes use of the explicit form of $f$.

The DSR-graph exists in different versions depending on the labelling of the edges (signs versus stoichiometric coefficients) and whether  two opposite directed edges between the same node pair are combined into one undirected edge or not \cite{banaji-craciun2,banaji-craciun1,wiuf-feliu}. 
It has been shown that the existence of positive feedback loops in the DSR-graph is a necessary condition for the system  \eqref{eq:ode2}  to admit multi-stationarity  \cite{banaji-craciun1}. 

Based on these results it has become standard to highlight positive feedback loops in multi-stationary reaction networks, eg. \cite{santos,Markevich-mapk}. The loops are typically found using intuitive reasoning that might overlook the existence of other relevant positive feedback loops or might select positive feedback loops that are not related to the existence of multi-stationarity. Here we  provide a method, based on theoretical considerations, to 
 classify \emph{all} positive feedback loops of a multi-stationary network into those that are related to the observed   multi-stationarity and those that are not.  In other words, we determine the positive feedback loops that when broken, multi-stationarity disappears. 
 
 The question needs to be understood in the context of the whole network and not in isolation: a particular positive feedback loop that is responsible for multi-stationarity in one network might appear in another network that cannot have multiple steady states. 

We present an automated procedure to determine the positive feedback loops that contribute to multi-stationarity. The procedure  is based on the \emph{injectivity} property applied to an  ODE system of the form \eqref{eq:ode2}, as described in  \cite{wiuf-feliu}. In this context, we review the proof of the fact that positive feedback loops are necessary for multi-stationarity and how this  relates to the DSR-graph. We illustrate the procedure with  examples of multi-stationary reaction networks involved in cell signaling. We further consider  the networks in the Biomodels  database \cite{BioModels2010} and apply the procedure to all non-injective networks (injective networks cannot be multi-stationary, see below). This provides an overview of the landscape of relevant positive feedback loops occurring in documented reaction networks.

\section*{Methods}
We use the formalism of  Chemical Reaction Network Theory (CRNT) \cite{feinbergnotes}. An ODE system is built from a set of reactions and reaction rates.

\subsection*{Reaction networks}
\begin{figure}[h!]
\center{\includegraphics{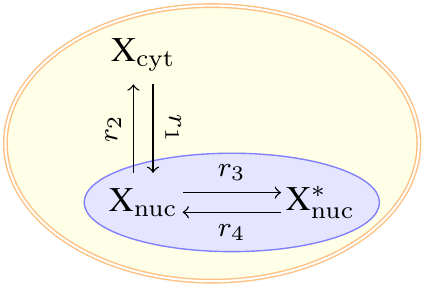}}
\caption{{\bf Main example.}  The reaction network used in  \cite{santos} as a toy model to  model the onset of mitosis. Here X is the complex Cdk1-cyclin B1 formed by the cyclin dependent kinase Cdk1 and the mitotic cyclin B1,  ``cyt'' indicates that the species is in the cytoplasm, ``nuc'' that it is in the nucleus, and X$^*$ is phosphorylated  CdC1-cyclin B1. Phosphorylation of Cdk1-cyclin B1 only takes place in the cell nucleus.
}\label{fig:santos}
\end{figure}

A \emph{reaction network}, or simply a \emph{network}, consists of a set of species $\{X_1,\dots,X_n\}$ and a set of reactions of the form: 
\begin{equation}\label{eq:reaction}
r_j\colon \sum_{i=1}^n \alpha_{ij} X_i \rightarrow \sum_{i=1}^n\beta_{ij} X_i, \qquad j=1,\dots,m
\end{equation} 
where $\alpha_{ij},\beta_{ij}$ are nonnegative integers, called the stoichiometric coefficients. As a running example we use the network in Figure  \ref{fig:santos}. It has three species, X$_{\text{cyt}}$, X$_{\text{nuc}}$, X$^*_{\text{nuc}}$, which are different forms of the Cdk1-cyclin B1 complex, and four reactions \cite{santos}. 

We denote the concentration of the species $X_1,\dots,X_n$ by lower-case letters $x_1,\dots,x_n$.  The evolution of the species concentrations with respect to time is modelled as an ODE system in the following way. We let $A=(a_{ij})$ be the stoichiometric matrix of the network:
\begin{equation*}
a_{ij} = \beta_{ij}-\alpha_{ij},
\end{equation*}
 that is, the $(i,j)$-th entry encodes the net production of species $X_i$ in reaction $r_j$. The vector $(a_{1j},\ldots,a_{nj})$ is called the \emph{reaction vector} of reaction $r_j$.
 
  The rate of reaction $r_j$ is a function $v_j\colon \Omega_v \rightarrow \R_{\geq 0}$, where $\R_{>0}^n\subseteq \Omega_v  \subseteq \R^n_{\geq 0}$ and $\Omega_v$ is the set of possible species concentrations. A typical choice of $v=(v_1,\dots,v_m)$ is \emph{mass-action kinetics}. In this case
\begin{equation*}
v_j(x)= x_1^{\alpha_{1j}}\cdot \dots \cdot x_n^{\alpha_{nj}},\quad x\in\Omega_v,
\end{equation*}
with the convention  that $0^0=1$.  Putting the pieces together provides a model for the evolution of the species concentrations over time:
\begin{equation}\label{eq:ode}
\dot{x} = Av(x),\qquad x\in \Omega_v.
\end{equation} 

Returning to  Figure \ref{fig:santos}, we let $x_{1}, x_2, x_3$ be the concentrations of X$_{\text{cyt}}$, X$_{\text{nuc}}$, X$^*_{\text{nuc}}$, respectively. Following \cite{santos}, one model of the network is:
\begin{align}
\dot{x}_1 & =  -\kappa_1 x_1 + \kappa_2 x_2 \nonumber \\
\dot{x}_2&  = \kappa_1 x_1 - \kappa_2 x_2 - \frac{x_2 (x_2+x_3)^4 }{K^4+(x_2+x_3)^4} + \kappa_4 x_3 \\
\dot{x}_3 & = \frac{x_2 (x_2+x_3)^4 }{K^4+(x_2 +x_3)^4} - \kappa_4 x_3,   \nonumber 
\end{align}
where $\kappa_1,\dots,\kappa_4,K>0$ are parameters. It takes the form \eqref{eq:ode} with
 \begin{equation}\label{eq:A}
 A=\left( \begin{array}{rrrr} -1 & 1 & 0 & 0 \\ 1 & -1 & -1 & 1 \\ 0 & 0 & 1 & -1   \end{array}  \right),
 \end{equation}
  \begin{equation}\label{eq:v1}
  v(x)=\left(\kappa_1 x_1,\kappa_2 x_2,\frac{x_2 (x_2+x_3)^4 }{K^4+(x_2+x_3)^4},\kappa_4 x_3\right),
  \end{equation}
and $\Omega_v=\R^n_{\ge 0}$.  Observe that the phosphorylation reaction $\text{X}_{\text{nuc}} \rightarrow \text{X}^*_{\text{nuc}}$ has a reaction rate that depends on both the concentration of the reactant X$_{\text{nuc}}$ and the concentration of the product X$^*_{\text{nuc}}$. We also consider an alternative  model in which the rate of X$_{\text{nuc}}$ phosphorylation  depends on $x_2$ only:
   \begin{equation}\label{eq:v2}
  v(x)=\left(\kappa_1 x_1,\kappa_2 x_2, \frac{x_2^5 }{K^4+x_2^4},\kappa_4 x_3\right).
  \end{equation} 
 This alternative model  is also consistent with the set of reactions in Figure \ref{fig:santos}, but the third reaction is now independent of the amount of X$^*_{\text{nuc}}$.

\subsection*{Multi-stationarity}
The specific form of \eqref{eq:ode} implies that the trajectories of the ODE system are confined to the so-called \emph{stoichiometric compatibility classes}:
\begin{equation*}
\mathcal C_0= (x_0+ \im(A)) \cap \Omega_v,
\end{equation*}
where $x_0=x(0)$ in $ \Omega_v$ is the initial condition. That is, the trajectories are restricted to the space spanned by the reaction vectors. Of particular interest is the (relative) \emph{interior} of $\mathcal C_0$, also called the \emph{positive stoichiometric compatibility class}, given by $\mathcal C_0\cap \R^n_{>0}$. Any trajectory that starts in $\mathcal C_0\cap \R^n_{>0}$, stays there, but might be attracted towards the boundary.

A reaction network is said to be \emph{multi-stationary} if there exist two distinct  steady states in a positive stoichiometric compatibility class (but not necessarily in all classes). Equivalently, if there exist distinct positive $x,y\in \R^n_{>0}$ such that $Av(x)=Av(y)=0$ and $x-y\in \im(A)$. A network with one positive steady state and one steady state at the boundary is therefore not multi-stationary in this terminology.

The reaction network  in Figure~\ref{fig:santos} is multi-stationarity for some choice of  parameters with the rate vector  in \eqref{eq:v1} \cite{santos}, but not with the rate vector  in \eqref{eq:v2} (which will be shown later).

\subsection*{Influence matrix}
The concept of a positive feedback loop is   associated with structural network properties  and  qualitative features of the reaction rates. Therefore, we assume some regularity on the reaction rates that we encode into  an abstract  symbolic matrix, called the \emph{influence matrix}.  A feedback loop does not depend on   specific parameters  or the specific functional form of the reaction rates.

To proceed, we assume that the function $v_j(x)$ is strictly monotone in each variable $x_i$ and  define the influence matrix $Z=(z_{ij})$ as 
\begin{equation*}
 z_{ij} = \begin{cases}  \gamma_{ij} & \textrm{ if }v_j(x)\textrm{ increases in }x_i\\
-\gamma_{ij} & \textrm{ if }v_j(x)\textrm{ decreases in }x_i \\
  0 & \textrm{ if }v_j(x)\textrm{ is constant in }x_i,
  \end{cases}
  \end{equation*}
where $\gamma_{ij}$ are symbolic variables.
    
 The influence matrices associated with the two reaction rate vectors in \eqref{eq:v1} and \eqref{eq:v2} are given by
\begin{equation}\label{eq:Z1}
Z_1=\left( \begin{array}{cccc} \gamma_{1,1} & 0 & 0 & 0  \\ 0 &  \gamma_{2,2} & \gamma_{2,3} & 0  \\ 0 & 0  & \gamma_{3,3} & \gamma_{3,4} \end{array}  \right),
\end{equation}
and
\begin{equation}\label{eq:Z2}
Z_2=\left( \begin{array}{cccc} \gamma_{1,1} & 0 & 0 & 0  \\ 0 &  \gamma_{2,2} & \gamma_{2,3} & 0  \\ 0 & 0  & 0& \gamma_{3,4} \end{array}  \right),
\end{equation}
respectively. All influences are zero or positive.

\subsection*{DSR-graph}

\begin{figure}[h!]
\center{\includegraphics{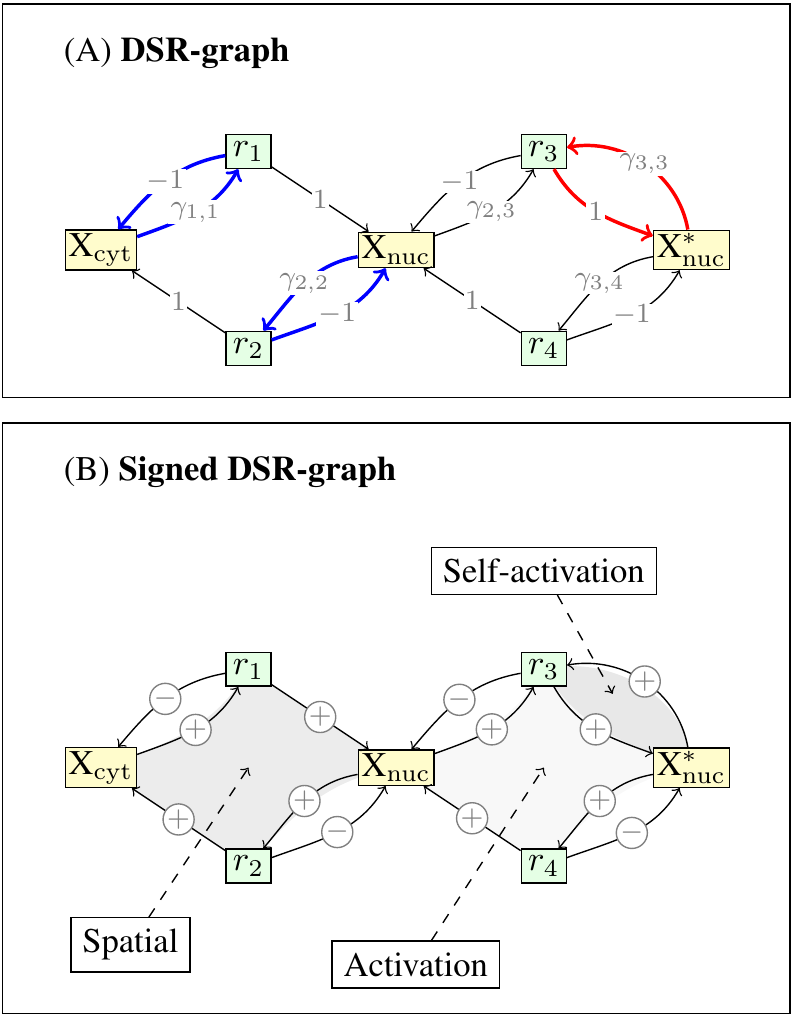}}
\caption{{\bf DSR-graphs of the running example.}  {\bf (A)} The DSR-graph. There are two 4-nuclei corresponding to negative terms in the polynomial $p_{A,Z_1}$: each of them consists of the red circuit combined with one of the two blue circuits. Of these, the only positive feedback loop is the red circuit, which is responsible for  the observed multi-stationarity.
{\bf (B)} There are three positive feedback loops in the graph, marked with shades of grey. Only the self-activation feedback loop (red circuit in (A)) is associated a term in the polynomial $p_{A,Z_1}$, see \eqref{eq:pAZ1}. Hence the other two positive feedback loops are not relevant for the observed multi-stationarity.}\label{fig:santosgraph}
\end{figure}

 We define the DSR-graph as a labelled bipartite directed graph with  node set $\{X_1,\dots,X_n,r_1,\dots,r_m\}$ and such that:
\begin{enumerate}[(a)]
\item There is an edge from $X_i$ to $r_j$ with label 
$z_{ij}$ if $z_{ij}\neq 0$.
\item There is an edge from $r_j$ to $X_i$ with label 
$a_{ij}$ if $a_{ij} \neq 0$.
\end{enumerate}
We refer to the \emph{signed DSR-graph} as the graph identical to the DSR-graph given by (a)-(b), but with the labels  replaced by their signs.  The (signed) DSR-graph of the running  example with $A$ as in \eqref{eq:A} and $Z$ as in \eqref{eq:Z1} is shown in Figure \ref{fig:santosgraph}.
The (signed) DSR-graph with $Z$ as in \eqref{eq:Z2} is identical to that in Figure \ref{fig:santosgraph}, with the edge from X$^*_{\text{nuc}}$ to $r_3$ removed.

\subsection*{Circuits and nuclei}
A \emph{circuit} in a graph $G$ is a sequence of distinct nodes $i_1,\ldots,i_q$ such that there is a directed edge from $i_k$ to $i_{k+1}$ for all $k\leq q-1$ and one from  $i_q$ to $i_1$.  A circuit must involve at least one edge.   
The label of a circuit $C$, denoted $\ell(C)$, is the product of the labels of the edges in the circuit. Two circuits are disjoint if they do not have any common nodes.

 A circuit with positive label is a \emph{positive feedback loop}, and similarly, a circuit with negative label is a negative feedback loop.
The three positive feedback loops of the running  example are shaded in Figure \ref{fig:santosgraph}B.  They  correspond to shuttling of the complex between the nucleus and the cytoplasm,  activation and deactivation of X$_{\text{nuc}}$, and  self-activation of X$_{\text{nuc}}$  (the rate of reaction $r_3$ increases with $x_3$, that is, the production of X$^*_{\text{nuc}}$ increases with the amount of X$^*_{\text{nuc}}$).

A $k$-nucleus  is a collection of disjoint circuits which involves $k$  nodes \cite{soule}.
 The label $\ell(D)$ of a  $k$-nucleus $D$ is the product of the labels of the edges in the nucleus.
 Let $a_1,a_2$ be the number of circuits in the nucleus that have odd (resp. even) number of species nodes and let $a=a_1+a_2$.
The sign of a $k$-nucleus is defined as  $\sigma(D)=(-1)^{a_2}$. 
That is, if  $D=C_1\cup \dots \cup C_a$  is  a disjoint union of circuits, then
\begin{equation}\label{eq:sD}
\sigma(D)\ell(D) =(-1)^{a_2} \prod_{i=1}^a  \ell(C_i).  
\end{equation}

In the DSR-graph,  any circuit involves an equal number of species and reaction nodes and, hence, an even number of edges. Consider a $2s$-nucleus $D=C_1\cup \dots \cup C_a$ of the DSR-graph. 
We show that if none of the circuits are positive feedback loops, then the sign of $\sigma(D)\ell(D)$ is $(-1)^s$. 
Indeed, if all circuits have negative labels, that is, $\ell(C_i)=-1$ for all $i$, then 
$$\sign(\sigma(D)\ell(D))  = (-1)^{a_2+a}  =  (-1)^{a_1+2a_2}  =  (-1)^{a_1}. $$
Because $D$ is a $2s$-nucleus, it contains $s$ species nodes. Let $n_i$ be the number of species nodes in circuit $C_i$, such that $s=n_1+\dots+n_a$. 
We have that $n_i$ is odd for $a_1$ of the circuits and even for $a_2$ of the circuits.
Therefore, $(-1)^{s} =   (-1)^{n_1+\dots+n_{a}} = (-1)^{a_1},$ and
\begin{equation}\label{signs}
\sigma(D)\ell(D)=(-1)^s
\end{equation}
 if there are no positive feedback loops  in $D$. This result is also in \cite{banaji-craciun2}, where it is stated using a different terminology.

\subsection*{Injectivity }

In this section we study injectivity of the function $x\mapsto Av(x)$, $x\in \mathcal C_0\cap\R^n_{>0}$  \eqref{eq:ode}. In the next section we link this injectivity property for all positive stoichiometric compatibility classes to the  non-existence of positive feedback loops in the DSR-graph. With other words, if all feedback loops are negative then the function is injective on all positive stoichiometric compatibility classes. In particular, there cannot exist two distinct points $x,y\in \R^n_{>0}$ in the same stoichiometric compatibility class such that $Av(x)=Av(y)=0$, that is, the network cannot be multi-stationary.

To decide whether the function $Av(x)$ is injective on all positive stoichiometric compatibility classes for any $v(x)$ with given influence matrix $Z$, we rely on a method previously developed by us \cite{Feliu-inj,wiuf-feliu,feliu-bioinfo}. We will now explain this method.

Given a pair of matrices $A,Z$, we define a polynomial $p_{A,Z}$ of degree  $s=\text{rank}(A)$, in as many variables as there are non-zero entries of $Z$. For this, let $M= A Z^t$ and let $\{\omega^1,\dots,\omega^d\}$ be a basis of  $\im(A)^\perp$, which we assume  to be Gauss-reduced. Further, let  $i_1,\dots,i_d$ be the indices of the first non-zero entries of $\omega^1,\dots,\omega^d$, respectively.  We define a  symbolic $n\times n$ matrix, $\nM$, by replacing the $i_j$-th row of $M$ with $\omega^j$  (cf. \cite[Section 5]{wiuf-feliu}). The polynomial $p_{A,Z}$ is defined as
\begin{equation*}
p_{A,Z}=\det(\nM),
\end{equation*}
which can be written as a  sum of  terms depending on the variables $\gamma_{ij}$, by expanding  the determinant. Each non-zero term is of the form $c\prod_{k=1}^s \gamma_{i_kj_k}$ where $c$ is a coefficient and  all $i_k$, respectively $j_k$, are distinct.

It is a result of \cite{wiuf-feliu,feliu-bioinfo} that if $p_{A,Z}$ is not identically zero and all non-zero coefficients of $p_{A,Z}$ have  the same sign, then the function $Av(x)$ is injective on each positive stoichiometric compatibility class and, hence, the network cannot be multi-stationary. As a consequence,   $p_{A,Z}$ has coefficients of opposite sign whenever the network is multi-stationary. If the coefficients do not have the same sign, then the network might be multi-stationary, but it cannot be concluded from the test.

Consider the matrix $A$  given in \eqref{eq:A} and $Z_1$ in \eqref{eq:Z1}.
We choose $\{(1,1,1)\}$ as a basis of $\im(A)^\perp$  and obtain
\begin{align}
 p_{A,Z_1} & =  -\gamma_{2,2}\gamma_{3,3}- \gamma_{1,1}\gamma_{3,3} \label{eq:pAZ1} 
 \\ &  \gamma_{2,2} \gamma_{3,4} + \gamma_{1,1}\gamma_{2,3}  + \gamma_{1,1}\gamma_{3,4}. \nonumber
 \end{align}
There are both positive and negative terms, hence multi-stationarity cannot be excluded. 
For $Z_2$ in \eqref{eq:Z2}, the polynomial $p_{A,Z_2}$ is obtained from \eqref{eq:pAZ1} by setting $\gamma_{3,3}=0$,
\begin{align}
 p_{A,Z_2} & = \gamma_{2,2} \gamma_{3,4} + \gamma_{1,1}\gamma_{2,3}  + \gamma_{1,1}\gamma_{3,4}. \label{eq:pAZ2} 
  \end{align}
 In this case all terms have the same sign and thus, the network cannot be multi-stationary. This holds for any choice of rate functions with influence matrix $Z_2$.

\subsection*{The polynomial and circuits}

In this section we link injectivity and the polynomial $p_{A,Z}$ to positive feedback loops.
It is shown in \cite{wiuf-feliu} that each term of the polynomial $p_{A,Z}$  can be identified with a  \emph{collection} of disjoint unions of circuits in the DSR-graph $G$. 
Specifically, given subsets $I,J\subseteq \{1,\dots,n\}$ of cardinality $s$,
let $D_s(I,J)$ be the set of $2s$-nuclei of $G$ with node set $\{X_i|\  i\in I\} \cup \{r_j |\ j\in J \}$. 
Then
\begin{equation}\label{eq:poly}
p_{A,Z} = \sum_{I,J\subseteq \{1,\dots,n\}} \sum_{D \in D_s(I,J)} \sigma(D)\ell(D), 
\end{equation}
where the sets $I,J$ in the sum have cardinality $s$ (cf. \cite[Section 11]{wiuf-feliu}). 

When $p_{A,Z}$ is derived from a ``reasonable'' reaction network, the predominant sign of the coefficients is  $(-1)^{s}$. This is because, for most reactions, the species in the reactant complex will have positive influence on the reaction rate, and the species in the product complex will have negative influence; if they influence the reaction at all. 
Consequently, if the network is multi-stationary, then  $p_{A,Z}$ has \emph{some} terms with the \emph{wrong} sign, that is, with sign $(-1)^{s+1}$.  Since   $\sign(\sigma(D)\ell(D))=(-1)^s$ whenever $D$ does not contain  positive feedback loops \eqref{signs}, we conclude that the network must contain positive feedback loops in order to be multi-stationary. 

Based on the above considerations, we  develop  a procedure to extract the positive feedback loops that correspond to terms with the wrong sign in $p_{A,Z}$. Fix a non-zero term of the polynomial $p_{A,Z}$, say
  \begin{equation}\label{term}
  (-1)^{s+1}c\, \gamma_{i_1,j_1}\dots \gamma_{i_s,j_s}
  \end{equation}
  ($c$ is positive) and consider the following edges from species to reactions 
\begin{equation*}
X_{i_k}\xrightarrow{\pm \gamma_{i_k,j_k}} r_{j_k}.
\end{equation*}
The $2s$-nuclei corresponding to the  term \eqref{term}  must contain these edges.
Therefore, we add to these edges all possible choices of $s$ edges from reactions $\{r_{j_1},\dots,r_{j_s}\}$ to species $\{X_{i_1},\dots,X_{i_s}\}$  such that the resulting graph is a $2s$-nucleus. We keep only the nuclei $D$ for which the sign of $\sigma(D)\ell(D) $ is  $(-1)^{s+1}$.
The positive feedback loops in these nuclei are those that \emph{do} contribute to the existence of multiple steady states.
Indeed, if all these loops are broken, then the network cannot be multi-stationarity. We find these loops in the signed DSR-graph.

For example, consider the polynomial $p_{A,Z_1}$ in \eqref{eq:pAZ1} and the DSR-graph shown in Figure~\ref{fig:santosgraph}.
In this case, the rank of $A$ is $s=2$, and hence we   focus on the negative terms since $(-1)^{s+1}=-1$. These are $\gamma_{2,2}\gamma_{3,3}$ and $\gamma_{1,1}\gamma_{3,3}$. The corresponding $4$-nuclei are depicted in Figure~\ref{fig:santosgraph}(A): there are two 4-nuclei obtained as the union of the red circuit and one of the two blue circuits. The only positive feedback loop that appears is therefore the self-activation positive feedback loop, and this is the only loop  that is related to the observed multi-stationarity. The other two positive feedback loops (termed the spatial and the activation loop, respectively in Figure \ref{fig:santosgraph}(B)) are therefore not relevant for the observed multi-stationarity.

\subsection*{Automated procedure}
The procedure to select positive feedback loops that contribute to multi-stationarity consists of the following steps.
\begin{enumerate}
\item For a network with stoichiometric matrix $A$ of rank $s$ and influence matrix $Z$, compute $p_{A,Z}$ and select the terms with sign $(-1)^{s+1}$.
\item Construct the DSR-graph. For each selected term of $p_{A,Z}$ with the wrong sign, determine the corresponding $2s$-nuclei   of the DSR-graph that have the wrong sign.
\item For each of the nuclei, select  the connected components that form positive feedback loops.
\end{enumerate}

 These steps have been implemented in Maple and the script is available upon request.  The procedure might fail for practical reasons (such as lack of computational memory) if the number of species and reactions is too big. In our experience, this number  depends heavily on the sparsity of the influence matrix \cite{feliu-bioinfo}.

\subsection*{Examples}
We have applied the procedure to several multi-stationary networks. 
These examples are also  in the Maple script, together with some other systems such as the three-site phosphorylation system.

\subsubsection*{Ring1B/Bmi1 ubiquitination system} 
We consider an  ODE model of histone H2A ubiquitination that  involves the E3 ligases Ring1B and Bmi1 \cite{Nguyen:2011p975}. When degradation of species is not taken into account, the  model has $10$ species and $15$ reactions.  We let  B and B$^d_{ub}$ denote the protein Bmi1 in isolation and targeted for degradation by ubiquitination, respectively.  The protein Ring1B is denoted by R, and R$_{ub}$, R$_{ub}^a$, R$^d_{ub}$ denote three different forms of self-ubiquitinated R, with R$^d_{ub}$ being the form targeted for degradation.  Bmi1 and Ring1B form a complex Z, that also might be  ubiquitinated, Z$_{ub}$.
Finally, Ring1B (either alone or in the complex Z) is responsible for the ubiquitination of the histone H2A, with  states H, H$_{ub}$.

The reactions describing the mechanism are \cite{Nguyen:2011p975}:
\begin{align*}
\text{B} & \cee{<=>[r_1][r_2]}\text{B}_{ub}^d  &  \text{R} & \cee{<=>[r_3][r_4]}  \text{R}_{ub}^d  &  \text{B}+\text{R} &  \cee{<=>[r_5][r_6]} \text{Z}  \\ 
  \text{Z} &  \cee{<=>[r_7][r_8]}  \text{Z}_{ub}   &      \text{Z}_{ub}  & \cee{<=>[r_9][r_{10}]} \text{B}+\text{R}_{ub}^a  & \text{R}  & \cee{<=>[r_{11}][r_{12}]} \text{R}_{ub} \\
   \text{R}_{ub}^a  & \cee{->[r_{13}]} \text{R}  & \text{H} &  \cee{<=>[r_{14}][r_{15}]} \text{H}_{ub}.
\end{align*}
We let  $x_1,\dots,x_{10}$ denote the concentrations of B, B$_{ub}^d$, R, R$_{ub}^d$, R$_{ub}$, R$_{ub}^a$, Z, Z$_{ub}$, H, H$_{ub}$, respectively.
The associated reaction rates are \cite{Nguyen:2011p975}:
\begin{align*}
v_1 &= \kappa_1x_1  & v_2 &= \kappa_2x_2   \\
v_3 & =\kappa_3x_3  & v_4 &=\kappa_4x_4   \\
 v_5  &=\kappa_5x_1x_3 & v_6 & =\kappa_6x_7 \\
v_7 &=x_7(\kappa_7x_7+\kappa_8x_8)   & v_8 &= \kappa_9x_8 /(\kappa_{10}+x_8 )  \\
v_9 &= \kappa_{11}x_8 &   v_{10}  &= \kappa_{12}x_1 x_6   \\
 v_{11} &=\kappa_{13}x_3^2+\kappa_{14}x_3x_5  & v_{12} &=\kappa_{14}x_5 \\
 v_{13}  &=\kappa_{15}x_6 &  v_{15} &=\kappa_{19}x_{10}, \\ 
 v_{14}  &=x_9 (\kappa_{16}x_5+ \kappa_{17}  x_8+ \kappa_{18}x_6),
 \end{align*}
where $\kappa_i>0$ are constants.

\begin{figure}[h!]
\center{\includegraphics[scale=0.72]{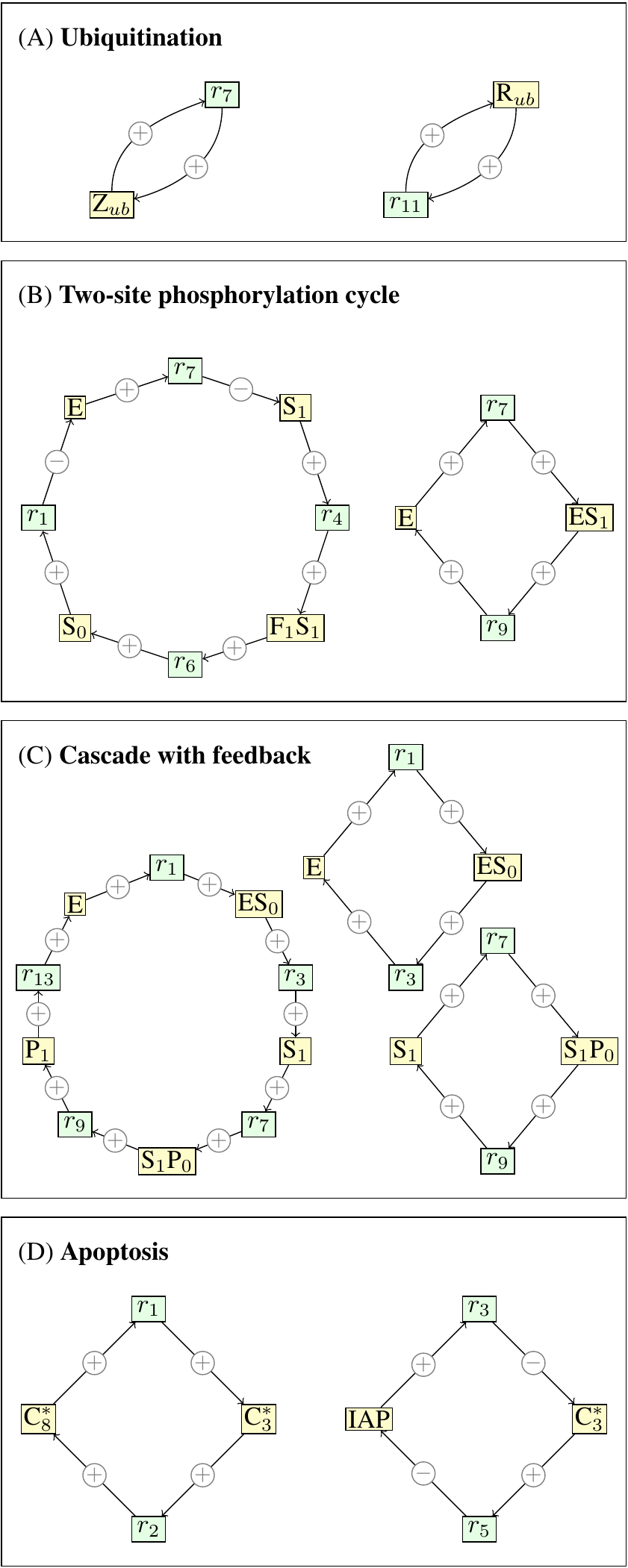}}
\caption{{\bf Examples.} 
{\bf (A)} For the ubiquitination system two  positive feedback loops are found. The loops correspond to self-ubiquitination of Z and R, respectively.
{\bf (B)} There are two positive feedback loops. The right loop corresponds to the Michaelis-Menten mechanism involving the two species E and ES$_1$. The left loop has four species nodes. The substrates S$_0$ and S$_1$ compete for the same kinase E in a way that enhances the production of both substrates: increasing S$_0$, decreases the amount of E (reaction $r_1$) which decreases the rate of reaction $r_7$, which in turn increases the amount of S$_1$.
{\bf (C)} Of the three positive feedback loops that are found, two correspond to the Michaelis-Menten mechanism (right side). One involves the kinase E and the complex ES$_0$. The second is similar, involving the kinase S$_1$ of the second layer and the complex S$_1$P$_0$. The left loop has five species nodes and illustrates P$_1$-activation of the kinase E.
{\bf (D)} The apoptosis system has two loops. The left loop occurs because C$_8^*$ in reaction $r_1$ increases the amount of C$_3^*$, which in turn increases the amount of C$_8^*$ via reaction $r_2$.
}\label{fig:examples}
\end{figure}

Self-ubiquitination of B is taken into account  in the rate functions $v_7$ and $v_{11}$ for reactions  $r_7$ and $r_{11}$, respectively. These incorporate a positive influence from the reaction products. With these specific rate functions,  the system  is multi-stationary \cite{Nguyen:2011p975}. We apply the automated procedure to find positive feedback loops and obtain the circuits depicted in Figure \ref{fig:examples}(A).  In \cite{Nguyen:2011p975}, it is postulated that  self-ubiquitination of Z and R are crucial steps for the emergence of multiple steady states, and we  confirm the statement here. 

\subsubsection*{Phosphorylation systems}
We have analysed different networks of signal transmission based on phosphorylation. We consider  models for sequentially distributed phosphorylation and dephosphorylation cycles and some modifications of these, see e.g.~\cite{Feliu:royal,G-distributivity}.

We consider first a two-site sequential phosphorylation cycle for a substrate S, where phosphorylation of the two sites is catalysed distributively by a kinase E, and  dephosphorylation of the two sites uses different phosphatases F$_1$, F$_2$. Assuming a Michaelis-Menten mechanism, the reaction network consists of the following reactions:
 \begin{align*}
\text{E}+ \text{S}_0  \cee{<=>[r_{1}][r_{2}]} \text{ES}_0   \cee{->[r_{3}]} \text{E} + \text{S}_1 \cee{<=>[r_{7}][r_{8}]} \text{ES}_1   \cee{->[r_{9}]} \text{E} + \text{S}_2 \\
\text{F}_1 + \text{S}_1 \cee{<=>[r_{4}][r_{5}]} \text{F}_1\text{S}_1   \cee{->[r_{6}]} \text{F}_1 + \text{S}_0  \\
\text{F}_2+ \text{S}_2  \cee{<=>[r_{10}][r_{11}]} \text{F}_2\text{S}_2  \cee{->[r_{12}]} \text{F}_2 + \text{S}_1 
\end{align*}
In S$_0$, S$_1$, S$_2$ the subindex denotes the number of phosphorylated sites. With mass-action kinetics, this system is  multi-stationary \cite{Markevich-mapk,Feliu:royal}. 
However, the  positive feedback loops that can  account for the observed multi-stationarity are not trivial.

We apply the automated procedure and obtain the positive feedback loops given in Figure~\ref{fig:examples}(B). The first of the two  loops has two  edges with negative labels. It  implies that S$_0$ and S$_1$ enhance their own production. Indeed, because S$_0$ and S$_1$ both compete for the same kinase, an increase in S$_0$  increases the  rate of reaction  $r_1$, which in turn lowers the amount of available enzyme E. This implies that reaction $r_7$ slows down and hence S$_1$ is consumed at a slower rate. The circuit closes through the reactions $r_4$ and $r_6$, which shows that an increase in S$_1$ implies an increase in S$_0$.

These type of loops are recurrent in  phosphorylation motifs. It is worth emphasising that the loops do not have independent meaning outside to network. Another network with the same positive feedback loop might not  be multi-stationarity.  This is apparent from the second positive feedback loop in Figure~\ref{fig:examples}(B), which is present whenever a Michaelis-Menten enzymatic mechanism is considered.

\subsubsection*{Signalling cascades}
We consider a 2-layer cascade with an explicit positive feedback. 
The first layer is a phosphorylation cycle with  kinase E,  phosphatase F$_1$, and phosphorylated and unphosphorylated substrate S$_0$, S$_1$. The second layer has kinase S$_1$,  phosphatase F$_2$, and phosphorylated and unphosphorylated substrate P$_0$, P$_1$.  Assuming a Michaelis-Menten mechanism, the reaction network consists of the following reactions:
\begin{align*}
\text{E}+ \text{S}_0  \cee{<=>[r_{1}][r_{2}]} \text{ES}_0    \cee{->[r_{3}]} \text{E} + \text{S}_1 \\
\text{F}_1 + \text{S}_1 \cee{<=>[r_{4}][r_{5}]} \text{F}_1\text{S}_1    \cee{->[r_{6}]} \text{F}_1 + \text{S}_0 \\
\text{S}_1+ \text{P}_0  \cee{<=>[r_{7}][r_{8}]} \text{S}_1\text{P}_0   \cee{->[r_{9}]} \text{S}_1 + \text{P}_1 \\ 
\text{F}_2 + \text{P}_1 \cee{<=>[r_{10}][r_{11}]} \text{F}_2\text{P}_1   \cee{->[r_{12}]} \text{F}_2 + \text{P}_0 \\
\text{P}_1    \cee{->[r_{13}]} \text{E}. 
\end{align*}
The automated procedure finds three positive feedback loops, as shown in  Figure~\ref{fig:examples}(C). The first  loop is  expected and  appears because the product of the second layer, P$_1$, activates the kinase of the first layer, E. The other two loops are similar to those in Figure~\ref{fig:examples}(C).

\subsubsection*{Apoptosis}
We finally consider  a basic  model of caspase activation for apoptosis  \cite{Eissing2004}:
{\small \begin{align*}
\text{C}_8^*+\text{C}_3 & \cee{->[r_1]} \text{C}_8^*+\text{C}_3^*  &  \text{C}_8 &\cee{<=>[r_{6}][r_{7}]} 0\\
\text{C}_8+\text{C}_3^* &  \cee{->[r_2]} \text{C}_8^*+\text{C}_3^* & \text{C}_3 &  \cee{<=>[r_{8}][r_{9}]} 0\\
 \text{C}_3^*+\text{IAP} &  \cee{<=>[r_{3}][r_{14}]}  \text{Y} \cee{->[r_4]} 0 & \text{IAP} &  \cee{<=>[r_{10}][r_{11}]} 0 \\
\text{C}_3^*+\text{IAP} & \cee{->[r_5]} \text{C}_3^*\cee{->[r_{13}]} 0   & \text{C}_8^* & \cee{->[r_{12}]} 0.
\end{align*}}
With mass-action kinetics, this network is  multi-stationary  \cite{Eissing2004} and has two relevant positive feedback loops,  Figure~\ref{fig:examples}(D). The second loop consists of two species, each with positive influence on a reaction rate, while at the same time, decreasing the amount of the other.

\subsection*{Analysis of the Biomodels database}

We  investigated the models in the database PoCaB \cite{samal2012}, which consists of 365 models from the  publicly available database Biomodels \cite{BioModels2010} (see also the page  http://www.ebi.ac.uk/biomodels-main/). The database PoCaB contains pre-computed stoichiometric matrices, mass-action exponent matrices, and kinetic data from the selected models.

 In a previous paper \cite{feliu-bioinfo} we extracted influence matrices of the reported  kinetics. Of the 365 networks, 323 have strictly monotone kinetics such that the influence matrix is well defined. On these 323 networks we ran the injectivity method and ended up with a total of 64 non-injective networks, excluding 27 very large networks where the injectivity method failed (as described in \cite{feliu-bioinfo}). Non-injectivity is a prerequisite for being multi-stationary. 

\begin{figure}[h!]
\center{\includegraphics{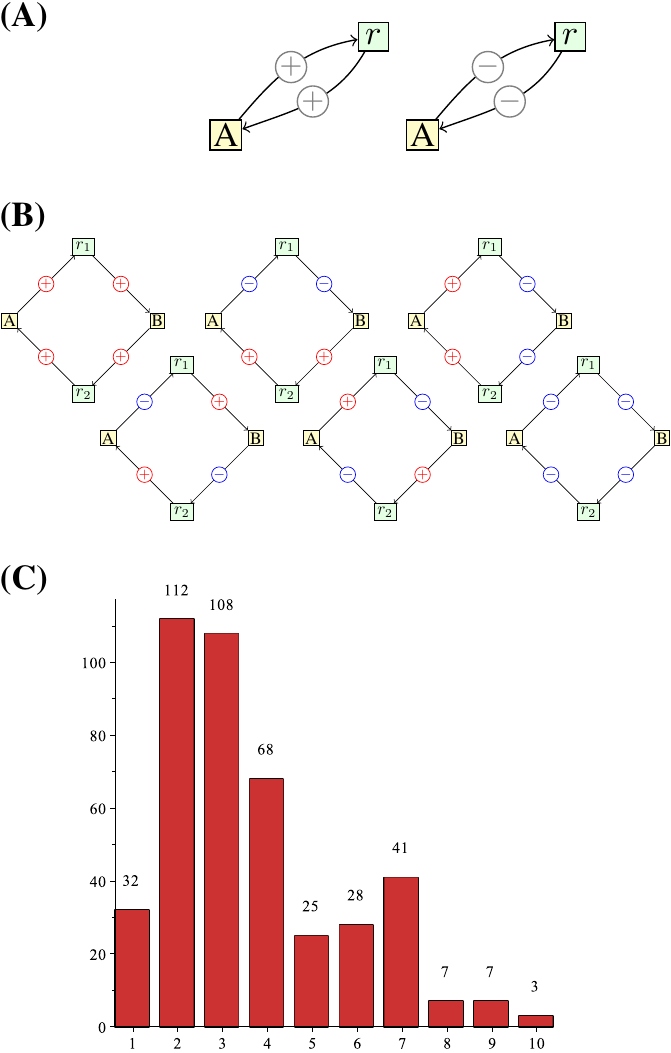}}
\caption{{\bf Biomodels database.} {\bf (A)} The positive feedback loops with one species. Among the 32 loops with one species, the frequencies are 19 and 13. {\bf (B)} The positive feedback loops with 2 species. Among the 108 loops with 2 species, the frequencies are (from top left, row by row): 35, 16, 9, 16, 13, 23.
{\bf (C)} The histogram shows the size (number of species) distribution among the 341 positive feedback loops found in the 64 models. 
}\label{fig:biomodels}
\end{figure}

We  applied the automated procedure on the 64 networks to determine the positive feedback loops. 
We obtained a total of 341 different positive feedback loops with size distribution shown in Figure~\ref{fig:biomodels}(C). Most   loops involve only 2 or 3 species (112 and 108 loops out of 341, respectively). In Figure~\ref{fig:biomodels}(A-B), we show  the positive feedback loops involving one or two species and conclude that all possible types occur in the database. However, for two species, their frequencies vary from 9 to 35 (out of 108),  indicating that the motifs are not equally represented in the database (Pearson's chi-square test, $p<0.0005$). For one species, there appears to be no difference ($p=0.28$).
We show in Table~\ref{table:biomodels} the most frequent positive feedback loops for different number of species nodes. 

\begin{table}[h!]
\renewcommand{\arraystretch}{1.1}
\center{\begin{tabular}{|clr|}
\hline
$N$ & Motif & Frequency\,\,\,\,\, \\ \hline
2 &  $(-, -, -, -)$ & 23/112 = 0.21\\
   & $(+, +, +, +)$ & 35/112 = 0.31 \\ \hline
3 & $(+, -, +, -, -, -)$ & 19/108 = 0.18 \\ 
   & $(+, +, +, +, +, +)$ &  18/108 = 0.17  \\\hline
4 & $(+, +,-, +, -, +, -, -)$ & 15/68 = 0.22\\ 
   & $(+, +, +, +, -, +, -, +)$ & 15/68 = 0.22\\ 
   & $(+, +, +, +, +, +, +, +)$  & 14/68 =  0.21 \\\hline
5 & $(+, +, +, +, +, +,-, +, -, +)$ & 9/25 =  0.36 \\ 
   & $(+, +, +, +, +, +, +, +, +, +)$ & 7/25 = 0.28 \\ \hline
6 & $( +, +, +, +,+, +, +, +, -, +, -, +)$  & 6/28 = 0.21\\ 
   & $(+, +, +, +, +, +, +, +, +, +, +, +)$ & 7/28 =  0.25 \\ \hline
7 & $(+, +, +, +, +, +, +, +, +, +, -, +,-, +)$ &  14/41 =  0.34 \\ \hline
\end{tabular}}
\caption{Positive feedback loops. For $N=2,\ldots,7$ species nodes, the most frequent ($>\!\!15\%$) positive feedback loops for each $N$ are shown, together with their frequencies.  At most four negative labels occur. Each cycle starts at a reaction node and the odd (even) labels correspond to reaction (species) nodes. Note that, for example, $(-,-,+,+,+,+)$ and $(+,+,-,-,+,+)$ are the same as they are permutations of each other. }\label{table:biomodels}
\end{table}

\section*{Discussion}

We have presented an automated procedure to find the positive feedback loops in a multi-stationary network.  The procedure relies on structural and qualitative features of the network together with a kinetics, and it is insensitive to the specific form of the rate functions.  Only positive feedback loops that are contributing to the multi-stationarity of the network are reported, excluding those positive feedback loops that are not relevant. 

Whether a loop is relevant or not, depends on the entire DSR-graph of the network (that is, the reactions and the influence matrix) and not just on the loop itself. In this sense, being a positive feedback loop related to an observed  multi-stationarity, is a context or network  dependent property.  This fact has also been observed in \cite{craciun-feinbergII,banaji-craciun1,banaji-craciun2}. In these papers, it  is shown that the existence of a positive feedback loop  fulfilling a certain extra  condition is a requirement for multi-stationarity to occur.
Specifically, the loop must either intersect another positive feedback loop in a specific way (called an \emph{S-to-R intersection}) or fulfil a certain condition on the labels  (called an \emph{s-cycle}). The first possibility is network dependent.  It is worth mentioning that there can be positive feedback loops that are s-cycles or that intersect another positive feedback loop in an S-to-R intersection without being relevant for the observed multi-stationarity. This is the case for most of the examples presented here. For example, the apoptosis network has another positive feedback loop, Y$\rightarrow r_{14} \rightarrow {\rm IAP} \rightarrow r_3 \rightarrow $Y (all edges are positive), which intersects the  loop on the right side in Figure~\ref{fig:examples}(D) in an S-to-R intersection.

The property of network dependence is further illustrated in Figure \ref{fig:santosgraph}(A)-(B), where the procedure is applied to the reaction network in Figure \ref{fig:santos} with influence matrix given by \eqref{eq:Z1}. The network  models translocation and phosphorylation of a cyclin dependent kinase X in the onset of mitosis. Only one of the three positive feedback loops shown in Figure \ref{fig:santosgraph}(B) can be associated with multi-stationarity in the model, namely the self-activating loop that  stimulates the creation of  phosphorylated X$^*_{\text{nuc}}$ in the nucleus.  In \cite{santos}, it is argued by different means than in this paper, that the spatial redistribution of the cyclin dependent kinase is important for creating the observed bistability. In contrast, our results suggest that the observed bistability is due to the self-activation loop of the phosphorylated complex in the nucleus.

The presented procedure cannot establish whether  a reaction network  is multi-stationary or not. Other means will here be required. If the procedure is applied to a reaction network  that might or might not be multi-stationary, then the absence of positive feedback loops  implies that the network cannot be multi-stationary, whereas the presence of positive feedback loops leaves the question open. 

Whether a positive feedback loop found by the procedure is important in a biological or experimental context, is not addressed in this paper, but must be established in other ways, for example by experimental verification. A  reaction network might only be multi-stationary for very specific choices of reaction rates, which might not be relevant in a particular experimental setting. As the procedure is parameter independent, any such verification and subsequent interpretation is beyond the scope of the procedure.

\section*{Acknowledgements}
CW was supported by the Danish Research Council, the Lundbeck Foundation (Denmark) and the Carlsberg Foundation (Denmark). EF was supported by project  MTM2012-38122-C03-01/FEDER from the Ministerio de Econom\'{\i}a y Competitividad (Spain), and the Lundbeck Foundation (Denmark). This paper was completed while the authors were visiting the Isaac Newton Institute in Cambridge, UK.

\end{document}